# Diagonal flipping of a Rhombus as elementary act of a polymorphic transformation. Calculation of energy threshold for the transformation in metals

**Michael Y Semenov** [a,*], **Valentin S Kraposhin** [a,**], **Alexander L Talis** [b],

[a]*Bauman Moscow State Technical University, 2nd Baumanskaya Street, 5, Moscow, 105005, Russian Federation,*

[b]*Laboratory of Polymer Physical Chemistry, A. N. Nesmeyanov Institute of Organoelement Compounds of the Russian, Academy of Sciences, Vavilova Street, 28, Moscow, 119991, Russian Federation, talishome@mail.ru,*

*Corresponding author: semenov.m.yu@bmstu.ru; **kraposhin@gmail.com.

**Abstract**

Diagonal flipping of a rhombus consisting of two triangles sharing a common edge with atoms (ions) in vertices of the triangles is considered as an elementary act of the polymorphic transformation in metals. The estimation of the energy threshold for the diagonal flipping has been carried out for various combination of rhombus vertices occupation by Fe, Cr, and Mn atoms. The energy threshold has been calculated in the framework of Morse interatomic pair potential. Numerical coefficients for the approximation of the pair potential function have been scaled by experimental values of the sublimation energy and temperature dependencies of elastic constants for Fe, Cr, Mn. Values of the energy threshold at 1193 K were estimated equal to 195, 150 and 94 kJ/mole for pure Cr, Fe and Mn respectively, i.e. in the same sequence as values of elastic bulk modulus for these metals. The substitution of one Fe atom by Cr or Mn atom results in alteration of the energy threshold. This alteration is dependent on the angle type at this vertex (an acute or obtuse angle), as well as on the ratio of bulk modulus values of iron and alloying element.

INTRODUCTION

The generally adopted crystallographic description of the polymorphic transformation FCC-BCC is usually reduced to the Bain scheme [1], or to the Kurdjumov-Sachs scheme [2-4]. Both schemes are represented as homogeneous deformation of the FCC lattice reconstructing it into the BCC lattice. The Bain deformation in combination with a rotation is equivalent to the two-shear Kurdjumov-Sachs deformation. However, this conventional approach cannot explain several significant experimentally observed geometrical features of this transformation, in particular, martensite habit plane (522)A is still out of explanation [4, 5] (here A means austenite). In our opinion, the limitations of the generally accepted approach are caused by the very essence of both schemes, namely by deformation of a lattice. Attempts to describe the crystal transformation by means of the lattice deformation (in most cases deformation of the unit cell are considered) is senseless from the physical point of view. Atoms are interacting with the next neighbors only, while the existence of an infinite crystal with a 3D-periodicity is ensured by the local Delaunay theorem only [6]. This theorem proves that the infinite translation lattice of any crystal is arising as a result of assembling of special finite clusters. And vice versa: each given structure is generated by a physical interaction of clusters, but not by a subdivision of the infinite lattice into parts. The concept of the unit cell is physically senseless, and this is especially clear in case of metals when atoms are occupying lattice vertices. In case of the crystal representation as sticking together of unit cells, one would be forced to cut atoms into parts [6]. The "lattice approach" of the Bain and Kurjumov-Sachs schemes is not corresponding to the local Delaunay theorem. Furthermore, the unit cell itself is ambiguously selected.

It is evident that accepted transformation models with lattice deformation cannot be improved anymore. By this reason we proposed an alternative approach in which the polymorphic transformation is described as the reconstruction of the clusters belonging to the initial structure into the clusters belonging to the resulting structure [8, 9]. The reconstruction of clusters is effected through the flipping of a rhombus diagonal; the mentioned rhombus is consisted from two neighboring triangles sharing common edge (Figure 1). The diagonal flipping in the single rhombus changes only its orientation on a plane. The result of the diagonal flipping in the 3D-structure is depicted in Figure 2. The diagonal flipping process is the Moebius symmetry transformation (a linear-fractional transformation), i.e. it is the transformation of a hyperbolic geometry not preserving a distance between two points, preserving the relation between coordinates of rhombus vertices [10-12]. This approach alone permits to describe all observed habite planes of the martensite in steels, {522)A habit plane being included [8], and crystal-geometric features of polymorphic transformation of Ti and Zr [13]. In this context the diagonal flipping corresponds to local theorem and does not contradict physics of an interatomic interaction. Contrarily to the diffusional jump by Zener, in case of the diagonal flipping all atoms remain in the same lattice vertices. Atom 3 is excluded from the next neighbors of atom 1, the rest next neighbors of atom 1 (atoms 2 and 4) are preserved. Atom 4 has an atom 2 as a new next neighbor. The diagonal flipping (switching of the interatomic bond) is similiar to the interatomic bond switching during the sliding of an edge dislocation (the displacement of an extra-plane).

The symmetry approach to the cluster transformation [14] allows to describe not only transformation between FCC, BCC, and HCP structures, but also the *in situ* transformation between $M_7C_3$ and $M_{23}C_6$ carbides in the heat resistant Fe-Cr-Ni-alloy [15]. Evidently, our local approach describes not the whole process of polymorphic transformation, but the first stage of the transformation: the formation of a nucleus of the new phase in the interior of the initial phase. The nucleus formation is strictly determined by the operations of a non-crystallographic symmetry. Experimentally observed crystal-geometry features of ferrous martensites have been successfully explained [16].

In the present paper the calculated estimation of the threshold energy for diagonal flipping in the rhombus was attempted (Figure 1). The calculations were made for various occupations of rhombus vertices: by atoms of the same element (Fe, Cr, or Mn), and also with substitution of Fe atoms in some vertices by Cr or Mn atoms.

ATOMISTIC MODELLING

In the initial state (Figure 1a) atoms positions correspond to the dense atomic packing of equal spheres on a plane with the interatomic spacing d=$2r_0$, here $r_0$ is the atomic radius. While calculating interatomic potentials the differences between atomic radii of iron $r_{Fe}$ and alloying element $r_X$ (X=Cr, Mn) have been taken into account for Fe-Cr and Fe-Mn clusters. During diagonal flipping the distance between atoms 1 and 3 increases, while the distance between atoms 2 and 4 decreases. Both spacings equal to $d\sqrt{2}$ at the saddle-point (Figure 1*d*), and at the flipping end (Figure 1e) the 1-3 distance increases up to $d\sqrt{3}$, while the 2-4 distance decreases from $d\sqrt{3}$ up to *d*. The obtained energy changes are summing up and this result is considered as energy threshold for the polymorphic transformation.

The energy threshold was calculated for the initial cluster configuration (the rhombus in Figure 1a) with several occupation variants of cluster vertices by different atoms (Figure 3). In the first case all vertices are occupied by the same atoms (Figure 1a). In the second (Figure 3) an alloying atom (Cr or Mn) substitute Fe atom in the vertex 1 of an obtuse angle of a rhombus, in the vertex 2 of the acute rhombus angle, in the rhombus vertices 1 and 4 (thus forming a rhombus edge), in vertices 1 and 3 (forming a short diagonal), in vertices 2 and 4 (forming a long diagonal of the rhombus).

In the framework of our symmetry approach for the description of structure transformation we suggest the geometric approximation of the conditionally hard spheres with an atomic compressibility which is characterized by an empiric coefficient λ. The atomic arrangement in a crystal is determined both by the achieving of a minimum of the Gibbs energy for a system, and by invariance of the system in the relation to one among 230 space symmetry groups. It should be noted that a space group is the closed set of matrices which is obtained from a purely geometric arguments without suggesting existence of both atoms and interatomic interactions. Thus the necessary conditions for atomic arrangement could be achieved by using symmetry invariants alone.

A type choice of the interatomic potential for calculations is always a subject for a speculation. The exponential approximations of the type of Lennard-Jones model [17] are frequently used:

$$U = 4\varepsilon\left[\left(\frac{2\sigma}{d}\right)^n - \left(\frac{2\sigma}{d}\right)^m\right] \quad (1),$$

where ε and σ are empirical coefficients; *n* and *m* −are exponents.

Here the values of power exponents and corresponding proportionality coefficients are selected with some degree of arbitrariness and could not be determined with the sufficient accuracy. In such a manner the values of n=8.5 and m=5.0 are recommended for iron [18]. Heinz et al [19] analyzed standard potentials with n/m 12/6 and 9/6, and more precise energetic and dimensional parameters were proposed to provide a good agreement between experimental and calculated data in relative narrow temperature range of 298±200 К.

The Pettifor approach [20] which was recommended for transition metals, we consider more adequate:

$$U = -\alpha\exp\left(-\frac{\lambda d}{2}\right) + \beta\exp(-\lambda d) \quad (2),$$

here α and β are n coefficients.

Potential (2) is the special case of the Morse potential [21] with using similar exponential relations. Morse approximation in its turn demands fitting of empirical or semi-empirical coefficients at exponential functions.

Calculations using the exponential Morse potential provide a reliable accordance with experiment as well, for example, see [22]. Lincoln *et al* [23] state that Morse potential shows maximal accordance to physical concepts of interatomic interactions. In the present paper we determine values of constant coefficients in equation (2) with calibration of the sublimation energy by experimental temperature dependence of shear modulus. Related mathematical transformations are presented in *Appendix A*.

Table 1 shows the initial atomic and electron structure data for Fe, Cr, and Mn which are needed to calibrations and calculations.

Table 1. Initial constants for calculations

| Element | Fe | Cr | Mn |
|---|---|---|---|
| Atomic compressibility $\lambda$, Å$^{-1}$ [24] | 3.2882 | 2.8158 | 3.0426 |
| Sublimation enthalpy $H_S$, kJ/mole | 415 [25] | 397 [26] | 272 [27] |
| Atomic diameter $d_0$, nm [28, 29] | 0.252 | 0.256 | 0.254 |
| Number of d-electrons $N_d$ | 6 | 5 | 5 |

RESULTS AND DISCUSSION

The primary checking of the model adequacy was fulfilled by means of calculating the predicted equilibrium value of the bulk elasticity modulus B. Relations between constants in the Morse-Pettifor equation (2) and the bulk modulus value B are derived in *Appendix B*.

Temperature dependence of the iron shear modulus G was taken from [30]. The corresponding regression curve $G = G(T)$ in Figure 4 and Table 1A was approximated with the confidence level of $R^2 = 0.912$.

Bulk elasticity modulus is temperature dependent [30]. The reliable experimental temperature dependence for iron bulk modulus in the vicinity of 1200 K is absent, hence the bulk modulus value was determined by known values of normal elasticity Young's modulus E and Poisson coefficient $\nu$. By [30] at T=1200 K for $\alpha$-Fe $E \approx 100$ GPa, for $\gamma$-Fe: $E \approx 120$ GPa.

Extrapolation of the averaged data of [30] up to 1200 K allows to obtain an estimation of $\nu \approx 0{,}35$. Thus, $B = 110\ldots130$ GPa. Calculation by these methods (Appendices A and B) gives B = 128.6 GPa for iron at 1193 K which is in good accordance with an experimental estimation.

Calculation errors are determined mainly by deviations of points of the G(T) curve from a straight line (see Figure 4), and furthermore by experimental errors in sublimation enthalpy values. The obtained regression relationships are valid above 500 K, below this temperature the linearity of the G(T) dependence is violated which became apparent as a smooth decrease of the Poisson coefficient from 0.35 up to 0.25-0.30 at 300 K [30].

By regression analysis of data in [31, 32] the $G = G(T)$ dependence for chromium metal has been obtained with the approximation confidence coefficient near to 1 (see Table 1A).

The determination accuracy for the constants in the pair Morse potential in case of chromium is significantly higher than in case of iron. Thus, in accordance with the calculation by adopted scheme (*Appendices A and B)* the bulk modulus B for Cr at 1193 K was found as equal to 173 GPa. Extrapolation of temperature dependencies of elastic constants for Cr [31] gives the estimation of bulk elasticity modulus as equal to 170 GPa.

The adequacy of the obtained equation for the Cr pair potential is limited from below at approximately 350 K. Below this temperature corresponding to Neel point magnetic phase transformation occurs in Cr and its elastic properties change significantly.

In case of Mn the pair potential ensures relatively high calculation accuracy for bulk elasticity modulus which is equal to 98 GPa at 973 K, while the extrapolation of experimental data gives a value of 102 GPa.

Table 2 shows calculation results for the energy threshold of diagonal flipping with several variants of the rhombus vertices occupation by Fe, Cr and Mn atoms (Figure 3).

Table 2. The energy threshold at two temperatures for several variants of iron atom substitution by chromium or manganese atoms at rhombus vertices

| The arrangement of Cr and Mn atoms (*) | Energy threshold (kJ/mole) at various temperatures | |
|---|---|---|
| | 1043 K | 1193 K |
| Fe atoms only | 180 | 150 |
| One Cr atom in the obtuse angle | 230 | 222 |
| One Cr atom in the acute angle | 131 | 93 |
| Cr atoms both in acute and obtuse angles (a rhombus edge) | 194 | 181 |
| Cr atoms in two obtuse angles (a short diagonal) | 250 | 240 |
| Cr atoms in two acute angles (a long diagonal) | 108 | 74 |
| Cr atoms only | 220 | 195 |
| One Mn atom in the obtuse angle | 146 | 126 |
| One Mn atom in the acute angle | 196 | 162 |
| Mn atoms both in acute and obtuse angles (a rhombus edge) | 150 | 126 |
| Mn atoms in two obtuse angles (a short diagonal) | 96 | 86 |
| Mn atoms in two acute angles (a long diagonal) | 206 | 170 |
| Mn atoms only | 110 | 94 |

(*) Iron atoms in the rest vertices

It could be seen that the placement of Cr atom into the obtuse angle of the cluster (vertex 1) increases the energy threshold of polymorphic transformation. The substitution of iron atom in the acute angle by chromium atom (vertex 2) decreases the threshold significantly. Such significant difference between energetic parameters of the polymorphic transformation with the various arrangement of chromium atom could be explained possibly by following speculations. In case of the placement of Cr atom into the obtuse angle one must stretch more rigid Fe-Cr atomic bonding along a short diagonal. Hence the resistance to stretching of the short diagonal is increased, and more energy is needed to overcome the transformation threshold. In case of the placement of Cr atom into the acute angle the energetic stimulus to move together Cr and Fe atoms along long diagonal is increased, so that the transformation threshold will be decreased. Evidently, in case of the placement of Cr atoms into two obtuse angles (short diagonal) one must stretch more rigid Cr-Cr-bonding than Cr-Fe bonding, hence the demanded energy for the transformation will be increased more significantly than in case of one Cr atom. Similarly, the placement of two Cr atoms in the long diagonal will give the maximal energy decrease if these Cr atoms will be forced move to each other. This case is similar to compressing of the most rigid string (Cr-Cr bonding) in comparing to the Cr-Fe-bonding and the weakest Fe-Fe bonding. The placement of Cr atoms in two acute angles ensures maximal decrease of the transformation threshold. It is possible that the interrelation between energy values with placements of Cr atoms in obtuse and acute angles will be apparent during the polymorphic transformation in a real 3-dimensional crystalline structure.

By introducing the Mn atom into the obtuse angle of the cluster (vertex 1) the energy threshold of polymorphic transformation decreased significantly due to weaker Fe-Mn-bonding in comparing to Fe-Fe-bonding. The substitution of Fe atom in the acute angle by Mn atom (vertex 2) results in the increase of the energy threshold for the same reason. In case of Mn atom in the obtuse angle one must stretch weaker Fe-Mn atomic bonding. Mn atom in the acute angle decreases the energy stimulus to move together Fe and Mn atoms; hence the total energy threshold for the transformation increases. The corresponding effects (an increase and decrease of the energy threshold) are observed while placing two Mn atoms onto long and short rhombus diagonal respectively.

Thus, the impacts of Mn and Cr atoms are opposite in sign. The magnitude of the Cr effect is approximately two times larger than Mn.

By application of the proposed calculation sequence one can predict the temperature dependence of the bulk modulus for the Fe-Cr alloy (Figure 5). It has been effected hereby with finite difference schedule while determining the second partial derivative of the pair potential in the equilibrium point.

As it was shown in Figure 5 the B(T) line for the Fe-Cr alloy is running approximately in the middle between the calculated lines corresponding to pure Fe and Cr, and also between experimental data for iron [30] and chromium [31]. In the literature experimental data for Cr above 1023 K are absent, due to this corresponding part of the curve has been obtained by extrapolation. The satisfactory accordance between calculated and experimental data could be seen. The maximal deviation for Fe at low temperatures could result from a nonlinearity of the temperature dependence of Poisson coefficient for Fe [30].

The calculated temperature dependence of the Mn bulk modulus is shown in Figure 6. One can see a satisfactory accordance with approximated experimental data. The B(T) curve for Fe-Mn alloy are running approximately in the middle between curves for pure Fe and Mn, similarly for above curves for Fe and Cr. It should be noted that temperature dependencies of the bulk moduli for Fe, Mn, and Fe-Mn alloy are approaching each other as the temperature is increasing but not diverging as in the case of Fe and Cr. It could be connected with the high temperature stability of the Cr bulk modulus.

The obtained data for the Fe-Cr cluster are supported by experimental results. Hillert [33] determined the energy activation for the migration of the α/γ interphase boundary in pure iron as equal to 147 kJ/mole in the total temperature interval of transformation from 550 up to 1250 K.

The energy activation for interphase boundary migration during the austenite-ferrite transformation has been estimated by Krielaart and Zwaag [34] as equal to 138 and 144 kJ/mole for steels containing 1 and 2%Mn respectively. Our calculations (Table 2) gave 146-150 kJ/mole with Mn atoms in acute and obtuse angles of rhombus. Experimental values of the energy of the austenite-ferrite transformation in the 0.005 % C, 9.6 % Cr steel at 1028 K and in the 0.007 % C, 9.3 % Cr steel at 1043 K has been obtained by magnetic method [35]. The obtained energy values were equal 199-212 and 212-231 kJ/mole respectively for two steels. Our calculations for cluster with two Cr atoms both in acute and obtuse angles resulted in the energy of 196 and 194 kJ/mole respectively at 1028 and 1043 K, which correspond sufficiently to the experiment.

The calculated values of the threshold energy for a transformation shows the increase of threshold value with the magnitude of the bulk modulus of a metal. Among three investigated metals chromium has the most bulk modulus value and is not undergoing a polymorphic transformation. Manganese with the lowest bulk modulus shows three polymorphic transformations in solid state. Melting points of three metals are close to each other, hence the stability of lattices relative the thermal elongation of interatomic bonds is almost the same for these three metals. Values of Young modulus, bulk modulus and melting points of several metals are shown in Table 3. These data allow finding a correlation of the metal capability to undergo a polymorphic transformation with the bulk elasticity modulus. Melting point of V is slightly higher the Ti melting point (Table 3) and normal Young moduli of two metals are differing on about 10%. Titanium metal undergoes a polymorphic transformation in solid state, but the vanadium metal with the bulk modulus on 45% higher than Ti bulk modulus has no solid state transformation. This effect is especially outstanding by comparing Pu and Al. Melting point of these two metals are almost the same, and Al bulk modulus is of 38% higher than Pu bulk modulus. Aluminum has no polymorphic transformations but in case of plutonium metal there are five solid state transformations (six polymorphic modifications).

The connection of the energy threshold with the value of the bulk modulus is in accordance with the estimation of a latent heat of melting for metals by the value of the bulk modulus [36]. The process of metal melting has been considered in [36] as the local transformation of the nearest coordination sphere of a close packed crystal lattice (a cuboctahedron) into the icosahedron. This well-known Kepler transformation could be presented as a combination of the considered here diagonal flipping in rhombi, forming the polyhedron with 12 vertices. Trivial calculation did show that the volume effect of the cuboctahedron-icosahedron transformation $\Delta V/V$ (a dilatation) was estimated as 16.6%, respectively the elastic energy of the cuboctahedron-icosahedron transformation could be estimated as $0.5(\Delta V/V)^2 B$. According to [36] for the majority of metals (Be metal being excluded only) this quantity is in good coincidence with the experimental values of the latent heat of melting. For example, elastic energy and heat of melting in case of Cu are equal respectively to 13.5 and 13.0 kJ/mole, for Al 10.4 and 10.67 kJ/mole, for Co 16.4 and 15.2 kJ/mole, for Ni 16.1 and 17.6 kJ/mole, for Fe 16.6 and 14.9 kJ/mole, for Na 2.8 and 2.64 kJ/mole, for K 2.5 and 2.40 kJ/mole. This approximate coincidence of the dilatation elastic energy and the latent heat of melting could be regarded also as the support for adequacy of the conditionally rigid sphere model which was used in the present paper.

Table 3. Properties of some metals manifesting or not manifesting polymorphism

| Element | Pu | Al | Ti | V |
|---|---|---|---|---|
| $E$, ГПа | 96 | 70 | 116 | 128 |
| $B$, ГПа | 55 | 76 | 110 | 160 |
| $\nu$ | 0.21 | 0.35 | 0.32 | 0.37 |
| Melting point, К | 912 | 934 | 1943 | 2160 |

Consideration of the diagonal flipping in a rhombus as an elementary act of the polymorphic transformation does not include the concept of a nucleus with the critical size. The energy fluctuation required for the diagonal flipping (to overcome an energy threshold could arise due to the difference between thermal expansions of different atomic bonds, i.e. due to an anisotropy of elongations in various crystallographic directions. The appearing elongation difference would violate the mechanical equilibrium of the mechanical system of the elastic rods which is similar to a space building construction. Each atom in this system plays the role of a hinge, ensuring a compliant bond between elastic rods. The further development of the transformation could be considered as the sequential flipping of the interatomic bonds in the neighboring cells by the "domino" principle up to the formation of a new crystal. The applied approach does not provide tool for the analysis of the transformation kinetics. We are considering the diagonal flipping only, which results in the formation of nanosized region with the new atomic coordination. The description of the overgrowth kinetics of this nanosized nucleus up to micro- and further to macro crystal is not considered in the framework of the used approach.

CONCLUSIONS

1. The flipping of the rhombus diagonal has been considered as an elementary act of the polymorphic transformations in metals. In the framework of the Morse pair potential the energy threshold for diagonal flipping with various arrangements of iron, chromium and manganese atoms in rhombus vertices has been estimated.

2. Constants in Morse pair potential expressions have been calibrated by energy sublimation values and experimental temperature dependencies of the shear modulus of iron, chromium and manganese metals. The calculation adequacy was supported by the coincidence of calculated bulk elastic moduli of iron, chromium, and manganese at given temperature with experimental values. The calculated values of the energy threshold of diagonal flipping in the iron-chromium cluster are also in the satisfactory accordance with the experimental values of the activation energy of the $\gamma \rightarrow \alpha$ transformation in Cr- and Mn- containing steels.

3. The magnitude of the energy threshold for diagonal flipping at 1193 K in rhombus with vertices occupied by atoms of one element only was calculated as equal to 195, 150, and 94 kJ/mole for Cr, Fe, and Mn respectively. Change of the energy threshold after substitution of one Fe atom by alloying element atom is dependent on the type of rhombus vertex (vertex of an acute or obtuse angle), and also on the ratio of bulk moduli for Fe and alloying element.

4. The placement of the atom of the element with bulk elastic modulus larger than bulk modulus of Fe (i.e. Cr atom) in the vertex of the obtuse angle results in the increase of the energy threshold from 150 up to 222 kJ/mole at 1193 K. In case of the substitution by atom of the element with lower bulk modulus (Mn atom) the decrease of the energy threshold down to 126 kJ/mole occurs. In case of the placement of the alloying atom in the vertex of the acute angle the interrelation between the energy threshold values is inversed: Cr atom decreases the threshold value down to 93 kJ/mole, while Mn atom increases the threshold value up 162 kJ/mole.

## APPENDIX A

According to Morse and Pettifor concepts it is required to develop expressions for U potential to each element [20], i.e. to solve a problem to find the pair potential for atoms of the single chemical element.

The equilibrium minimum $U_0$ of the interatomic potential at the equilibrium interatomic spacing $d = 2r_0$, could be set equal to the sublimation energy of metal:

$$U_0 = -U_s. \qquad (1A)$$

From the expression presented in [20] it follows that:

$$U = U_b + U_r = -\alpha \exp\left(-\frac{\lambda d}{2}\right) + \beta \exp(-\lambda d), \qquad (2A)$$

where $U_b$ – the energy of the mutual attraction of atoms (the electron bond energy), $U_r$ – the energy of the mutual repulsion (interaction energy for identically charged ion cores and electrons).

In the point of energy minimum at $d = 2r_0$ the first derivative of $U_0$ by interatomic distance is equal to zero, hence we obtain the (3A) expression, which in combination with the (2A) expression permits to determine α and β constants:

$$\frac{\partial U(d = d_0)}{\partial d} = \alpha\lambda\exp\left(-\frac{\lambda d_0}{2}\right) - 2\beta\lambda\exp(-\lambda d_0) = 0, \quad (3A)$$

where $d_0$ – is the equilibrium interatomic distance, $d_0 \approx 2r_0$.

Then:

$$\alpha = \frac{2U_0}{\exp(-\frac{\lambda d_0}{2})}, \quad \beta = \frac{U_0}{\exp(-\lambda d_0)}. \quad (4A)$$

α and β constants are temperature dependent through $U_0$.

Temperature dependence of the sublimation energy is expressed by the sum:

$$U_s = H_s + S_s T; \quad (5A)$$

where $H_s$ – sublimation enthalpy; $S_s$ – sublimation entropy at 450-1250 K; $T$ - temperature. $S_s$ was determined similarly to diffusion activation entropy by the formula in [37, 38]:

$$S_s = \frac{H_s}{G_0} \cdot \frac{\partial G}{\partial T}; \quad (6A)$$

here $G$ – shear modulus; $G_0$ – shear modulus at $T = 0$ K.

The obtained temperature dependencies of $G$ and $U_S$ are shown in Table 1A.

Table 1A. Regression expressions for temperature dependencies of shear modulus and sublimation energy

| Element | Fe | Cr | Mn |
|---|---|---|---|
| Shear modulus, GPa | $G = -0.0453T + 94.403$ | $G = -0.0225T + 121.75$ | $G = -0.0420T + 93.56$ |
| Sublimation energy, kJ/mole | $U_s = 415 - 0.2063T$ | $U_s = 397 - 0.0734T$ | $U_s = 272 - 0.1221T$ |

Thus, one can calibrate approximation constants α and β in the symmetrical pair potential by the temperature dependencies of elastic constants.

The obtained the mutual attraction and repulsion energies values for Fe and alloying element atoms were used to solve an asymmetric problem, i.e. to calculate mutual attraction and repulsion energies for Fe and X atom pairs, respectively, $U^{Fe-X}_b(d)$ and $U^{Fe-X}_r(d)$ at an arbitrary interatomic distance $d$ in accordance with expressions in [20]:

$$U_b^{Fe-X}(d) = -W^{Fe-X}(d) N_d^{Fe-X} \left(10 - N_d^{Fe-X}\right)/20, \quad (7A)$$

where $N^{Fe-X}_d$ – d-electrons mean number for iron and alloying element atoms; $W^{Fe-X}(d)$ – mean width of the valence zone for Fe and alloying element atoms at interatomic distance $d$:

$$W^{Fe-X}(d) = \left\{0.5\left[W^{Fe}(d)^2 + W^X(d)^2\right]\right\}^{\frac{1}{2}}, \quad (8A)$$

where $W^{Fe}(d)$ and $W^X(d)$, and further– $W^{Me}(d)$ – width of the valence zone for Fe and alloying element atoms at interatomic distance $d$:

$$W^{Me}(d) = \frac{20}{N_d} U_b^{Me}(d)(1 - N_d), \quad (9A)$$

and where $U^{Me}_b(d)$ – the mutual attraction energy calculated above for a given metal atom Me at distance $d$; $N_d$ – number of d-electrons;

$$U_r^{Fe-X}(d) = W^{Fe-X}(d) N_d^{Fe-X}\left(10 - N_d^{Fe-X}\right)/40, \quad (10A)$$

where $N^{Fe-X}_d$ – the arithmetic mean of electron numbers for Fe atoms and alloying element X, $N_d$.

Then interatomic pair potential for the cluster containing iron and alloying element at interatomic distance d can be determined as [20]:

$$U^{Fe-X}(d) = U_b^{Fe-X}(d) + U_r^{Fe-X}(d). \quad (11A)$$

To determine the temperature dependence of the bulk modulus for Mn we were forced to use an artificial method since there are no direct published data for it. From the other hand there is the temperature dependence of the Mn normal elasticity modulus [39]. The linear proportionality coefficient $\Delta_B$ is also known. This coefficient characterizes the influence of temperature on the Mn bulk modulus, and it is approximately equal to 0.22 [40].

By using this $\Delta_B$, value, and values of the Mn bulk modulus at room temperature (120 GPa), and melting point 1519 K, and suggesting the linear B(T) dependence we substitute the expression in [41]:

$$\frac{\Delta B}{\Delta T} = \Delta_B \frac{B(0\,K)}{B(1519\,K)}, \qquad (12A)$$

by the approximate relation:

$$\Delta_B \frac{B(0\,K)}{B(1519\,K)} \approx \Delta_B \frac{B(300\,K)}{B(1519\,K)} \times \frac{1519-0}{1519-300}. \qquad (13A)$$

The approximated B(T) dependence has been obtained as a result:

$$B = -0.0177T + 124.77. \qquad (14A)$$

Bearing in mind equation (13A), the known $E=E(T$ dependence [39] and using formula (15A) below:

$$G = \frac{3BE}{9B-E} \qquad (15A),$$

we obtained the regression relations for the shear modulus and sublimation energy calculation (see Table 1A).

The developed mathematical model based on atomistic calculations has been realized on the Object Pascal language in the original program permitting the calculations the dependence of the pair interaction energy for Fe, Cr, and Mn on the interatomic distance at an arbitrary temperature (Figure 7). Dependencies of the pair interaction energy for Fe and X atoms are not qualitatively differing from the corresponding dependencies for pure Fe. In case of different occupation of the opposite cluster vertices (see Figure 3) by Fe and X atoms we used (10A) formula for energy calculation, in case of single atoms arrangement formula (2A) was used.

APPENDIX B

It is known that at equilibrium interatomic distance, pressure and temperature being constant, the bulk elasticity modulus is equal [for example, 24, 42] to:

$$B(r = r_0) = \Omega \frac{\partial^2 U}{\partial \Omega^2}, \qquad (1B)$$

where $\Omega$ - atomic volume. The expression (1B) was converted with the determination of $\Omega$ by $r$:

$$B(r = r_0) = \frac{1}{12\pi r_0} \frac{\partial^2 U(r = r_0)}{\partial r^2}, \qquad (2B)$$

From (2B) after double differentiation and taking into account the expression for Morse potential, we obtained:

$$B(r = r_0) = \frac{\lambda^2}{12\pi r_0} \times [-\alpha \exp(-\lambda r_0) + 4\beta \exp(-2\lambda r_0)]. \qquad (3B)$$

FIGURE CAPTION:

Figure 1. The elementary act of the polymorphic transformation in metals is effected by the mutual substitution of long by short diagonal in a rhombus, and vice versa from (a) up to (e). Atoms in vertices 1 and 3 are moving apart, while in vertices 2 and 4 are moving together. In the saddle-point (d) diagonal lengths became equal.

Figure 2. Diagonal flipping in 3-dimensional space: flipping of the common 1-5 edge shared by three tetrahedra transforms three tetrahedra into an octahedron (the 4-fold axis appeared).

Figure 3. Initial configurations of the rhombus with different vertices occupied by alloying atoms, X=Cr or Mn.

Figure 4. Temperature dependence of the Fe shear modulus obtained by the approximation of experimental data from [29] (○).

Figure 5. Temperature dependencies of the bulk elastic moduli of Fe, Cr, and Fe-Cr alloy obtained by calculations (solid lines) and by experiment extrapolation (dotted lines).

Figure 6. Temperature dependencies of the bulk elastic moduli of Fe, Mn, and Fe-Mn alloy obtained by calculations (solid lines) and by experiment extrapolation (dotted lines).

Figure 7. The calculated dependence of the energy of the pair interaction on the interatomic distance for iron and chromium atoms at 1193 K. 1 – a repulsion energy $U_r(d)$, 2 – an attraction energy $U_b(d)$, 3 – a total bonding energy $U(d)$.

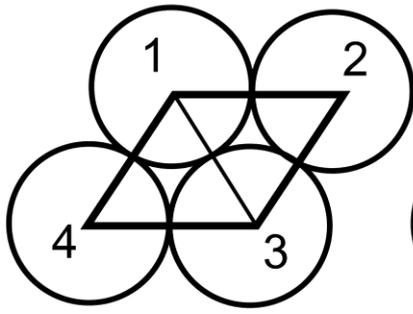
a)

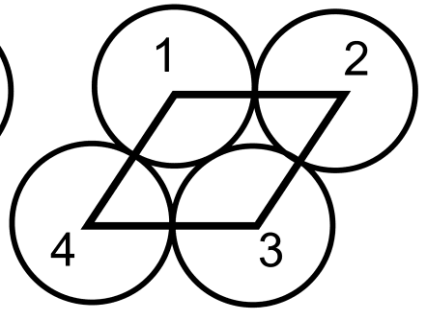
b)

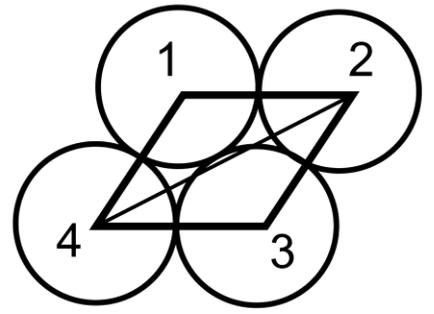
c)

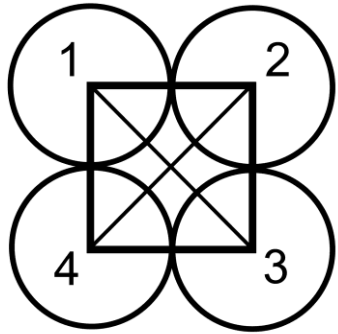
d)

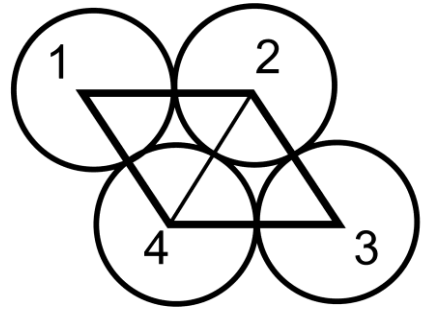
e)

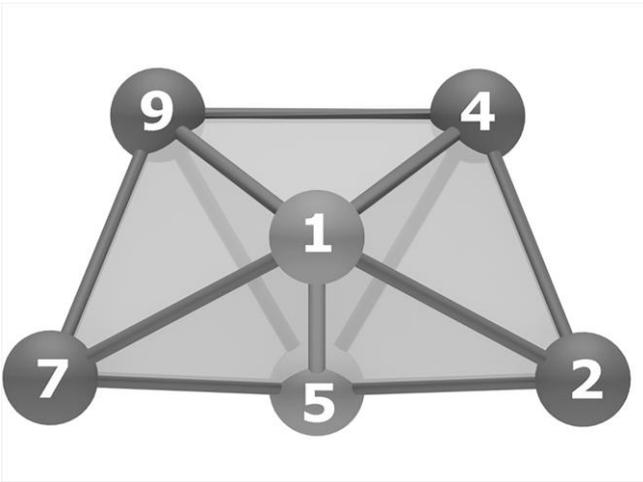

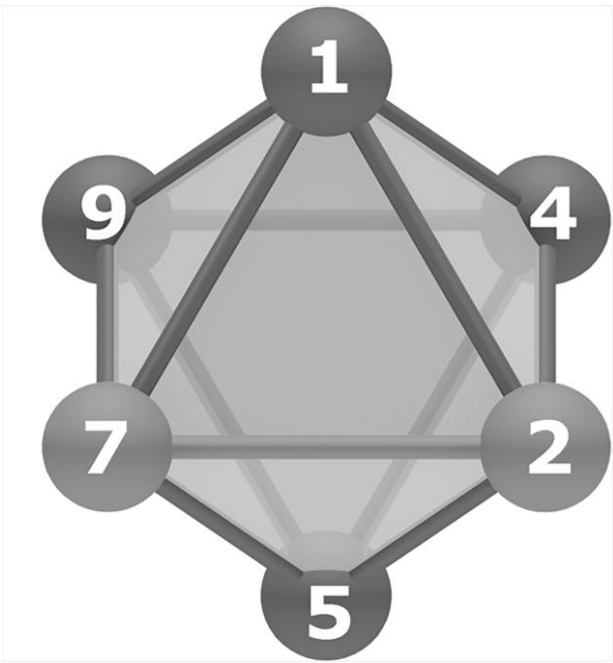

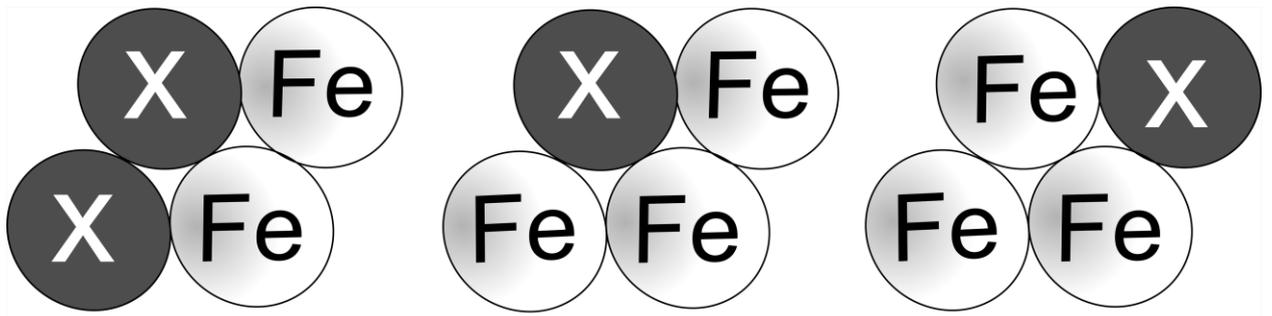

a) b) c)

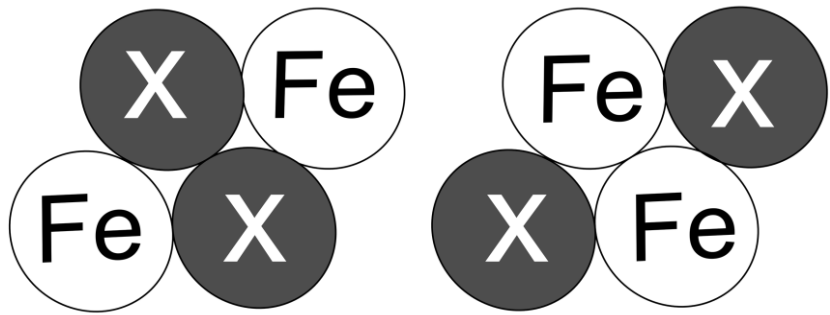

d) e)

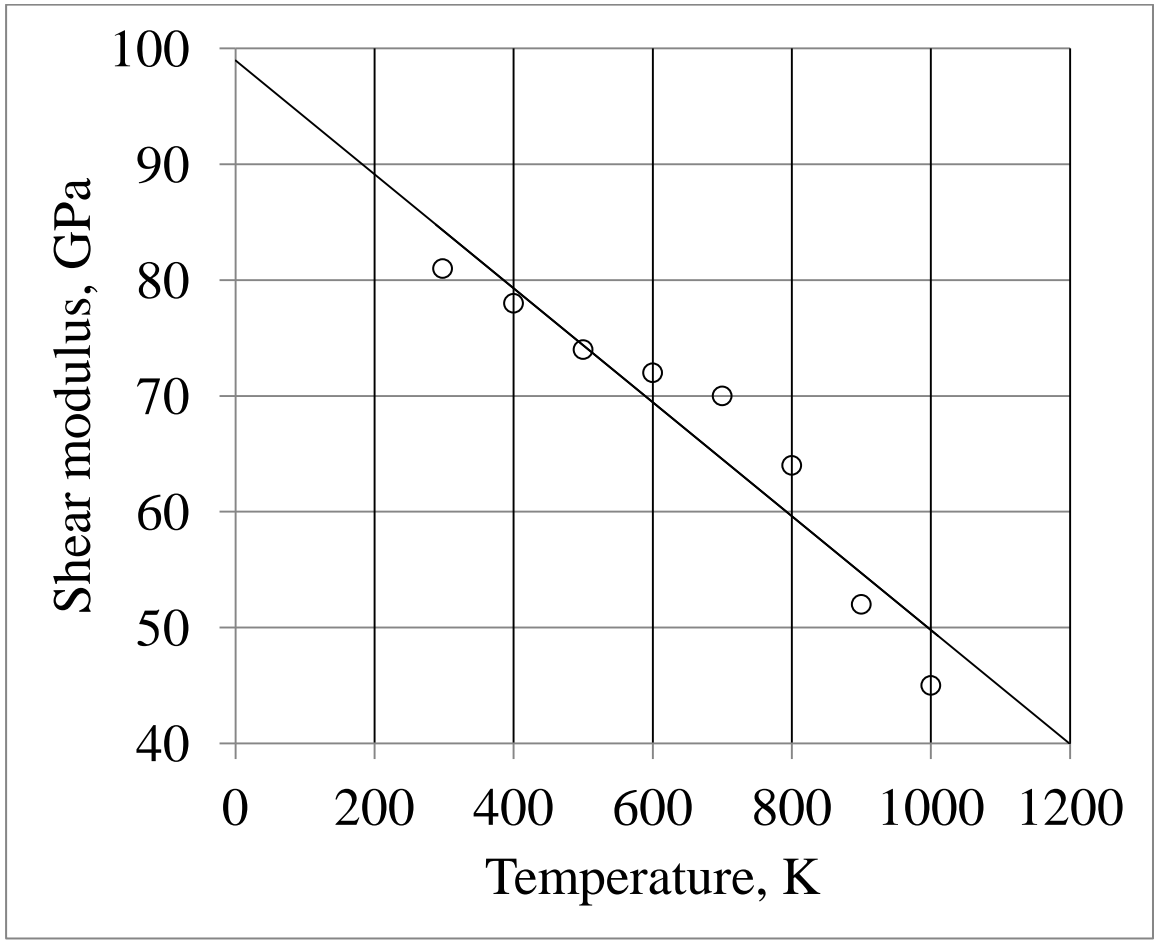

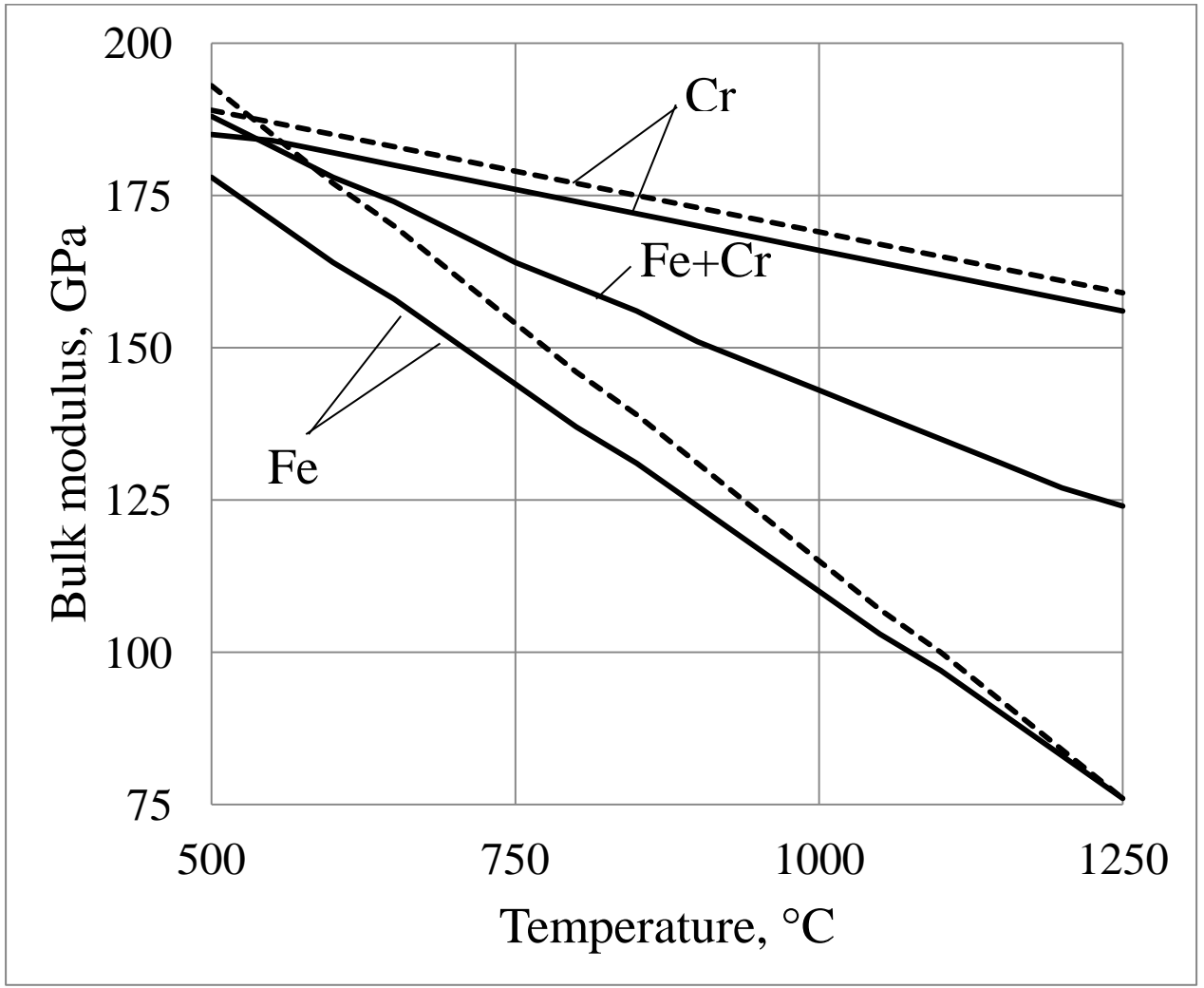

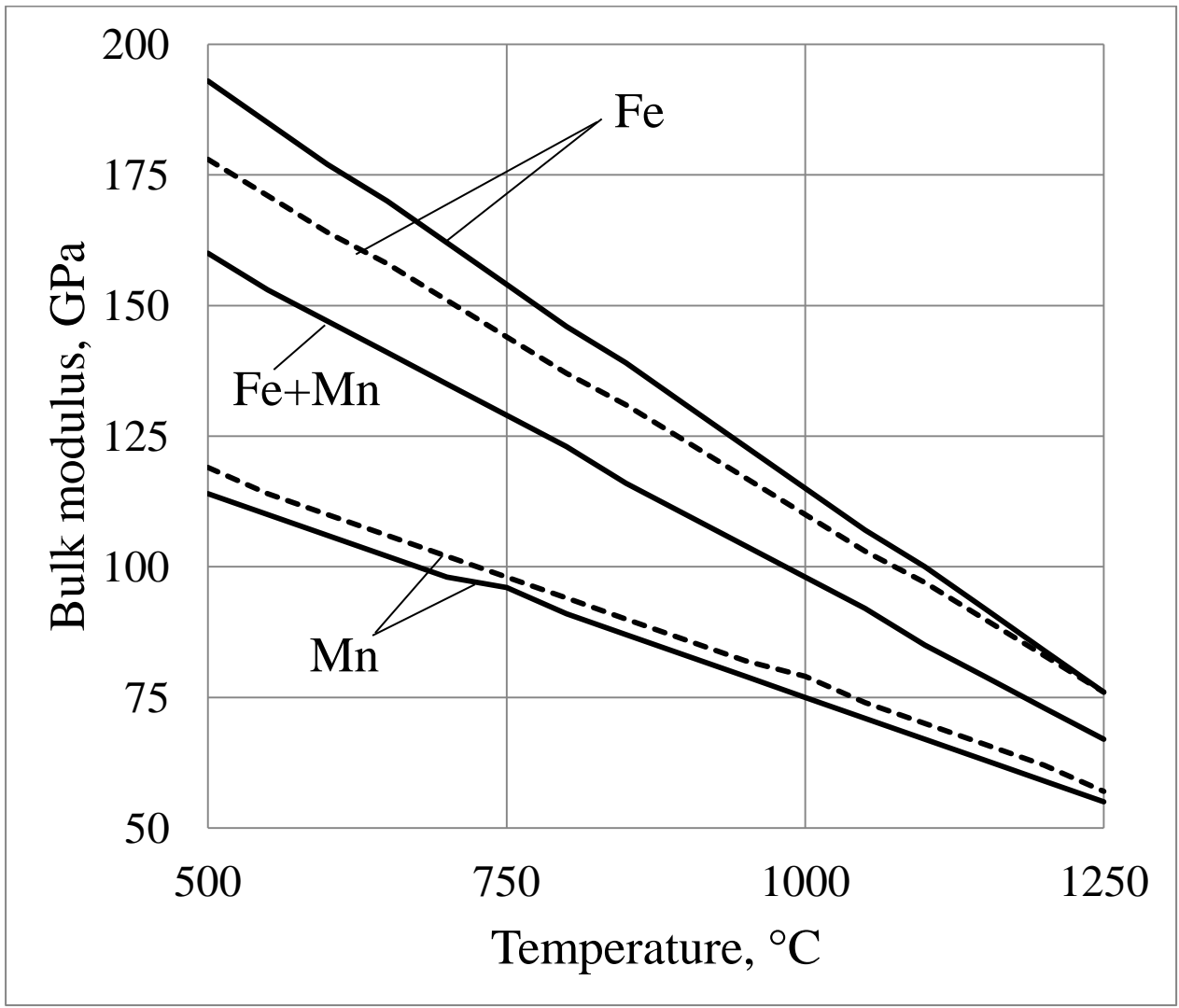

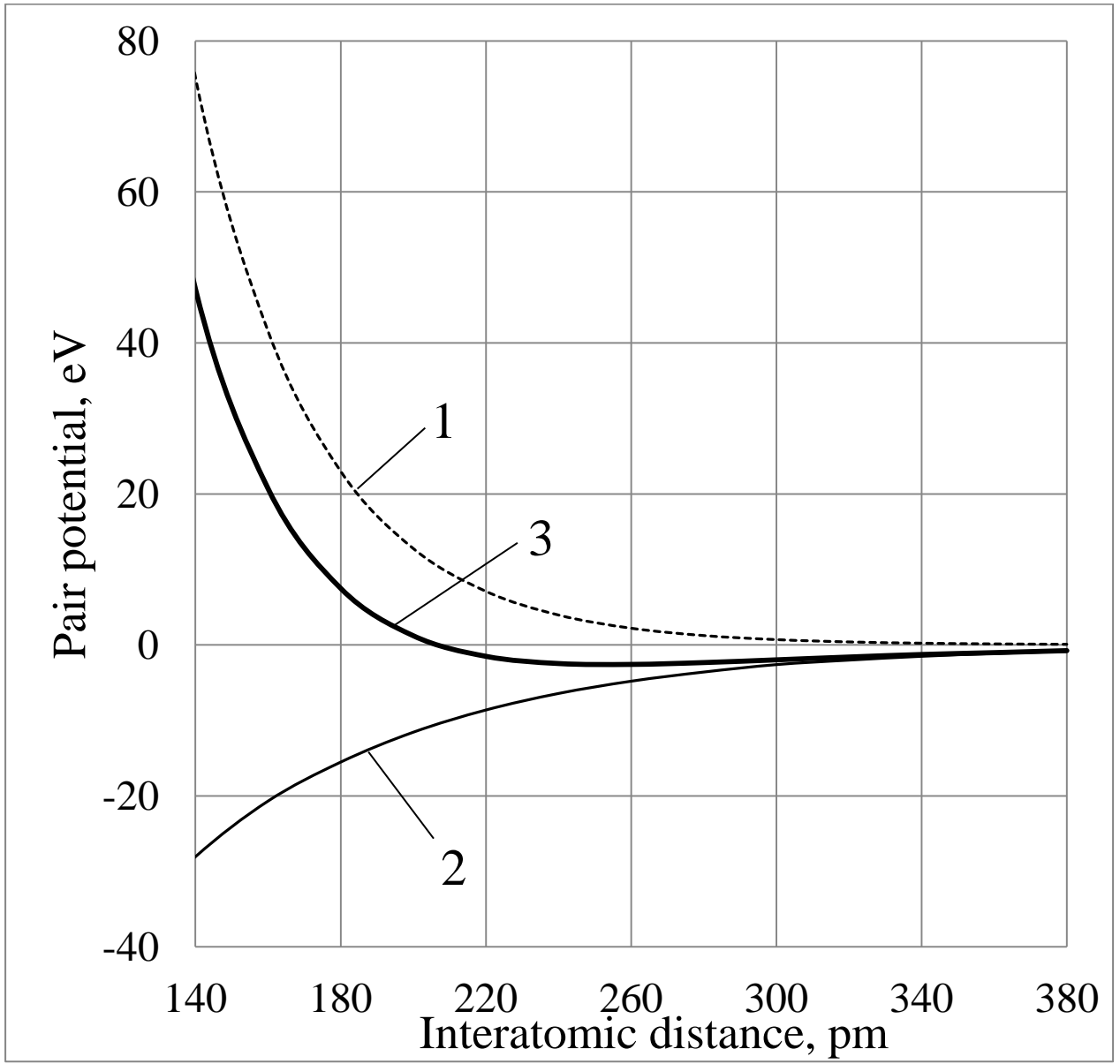